\documentclass[prl,10pt,twocolumn,superscriptaddress,aps,floatfix]{revtex4-1}
\usepackage{graphicx}
\usepackage{natbib}
\usepackage{url}
\usepackage[normalem]{ulem}
\usepackage{color}
\usepackage{mathtools,amssymb}
\usepackage{textcomp}
\usepackage{gensymb}
\usepackage{braket}
\newcommand{\dVdI}{\mathrm d V / \mathrm d I}

\begin{document}

\title{Andreev interference in the surface accumulation layer of half-shell InAsSb/Al hybrid nanowires}

\author{Lukas Stampfer}
\affiliation{%
Center for Quantum Devices, Niels Bohr Institute, University of Copenhagen, Universitetsparken 5, 2100 Copenhagen, Denmark.
}%

\author{Damon J. Carrad}

\affiliation{%
Center for Quantum Devices, Niels Bohr Institute, University of Copenhagen, Universitetsparken 5, 2100 Copenhagen, Denmark.
}%

\author{Dags Olsteins}%
\affiliation{%
Center for Quantum Devices, Niels Bohr Institute, University of Copenhagen, Universitetsparken 5, 2100 Copenhagen, Denmark.
}%

\author{Christian E. N. Petersen}%
\affiliation{%
Center for Quantum Devices, Niels Bohr Institute, University of Copenhagen, Universitetsparken 5, 2100 Copenhagen, Denmark.
}%

\author{Sabbir A. Khan}%

\affiliation{%
 Center for Quantum Devices, Niels Bohr Institute, University of Copenhagen, Universitetsparken 5, 2100 Copenhagen, Denmark.
}%
\affiliation{Microsoft Quantum Materials Lab Copenhagen, 2800 Lyngby, Denmark}

\author{Peter Krogstrup}%
\affiliation{%
 Center for Quantum Devices, Niels Bohr Institute, University of Copenhagen, Universitetsparken 5, 2100 Copenhagen, Denmark.
}%
\affiliation{Microsoft Quantum Materials Lab Copenhagen, 2800 Lyngby, Denmark}

\author{Thomas S. Jespersen}%
\email{tsand@nbi.ku.dk}
\affiliation{%
 Center for Quantum Devices, Niels Bohr Institute, University of Copenhagen, Universitetsparken 5, 2100 Copenhagen, Denmark.
}%
\affiliation{Department of Energy Conversion and Storage, Technical University of Denmark, Fysikvej, Building 310, 2800 Kgs. Lyngby}

\begin{abstract}

Understanding the spatial distribution of charge carriers in III-V nanowires proximity coupled to superconductors is important for the design and interpretation of experiments based on hybrid quantum devices. In this letter, the gate-dependent surface accumulation layer of InAsSb/Al nanowires was studied by means of Andreev interference in a parallel magnetic field. Both uniform hybrid nanowires and devices featuring a short Josephson junction fabricated by shadow lithography, exhibited periodic modulation of the switching current. The period corresponds to a flux quantum through the nanowire diameter and is consistent with Andreev bound states occupying a cylindrical surface accumulation layer. The spatial distribution was tunable by a gate potential as expected from electrostatic models. 

\end{abstract}

\maketitle
Semiconductor nanowires (NWs) with superconducting electrodes provide a flexible experimental  platform for studies of mesoscopic superconductivity. Experiments have demonstrated tunable\cite{Doh:2005} and reversible\cite{vanDam:2006} supercurrents, Cooper pair splitting\cite{Hofstetter:2009}, electron pumping \cite{Giazotto:2011}, gate-tunable superconducting qubits\cite{Larsen:2015}, and have provided a detailed understanding of Andreev bounds states (ABS)\cite{Prada:2020} and topological Majorana zero modes\cite{Oreg:2010,Lutchyn:2010,Mourik:2012,Albrecht:2016}.
The performance of NW hybrid devices is strongly affected by crystal defects\cite{Kretinin:2010,Schroer:2010}, interface disorder\cite{Krogstrup:2015}, and surface scattering\cite{Lynall:2018} and the experimental progress has gone hand-in-hand with developments of materials and methods for device fabrication to minimize sources of disorder. High-mobility semiconductor NWs can now be grown into complex geometries\cite{Dalacu2013,gazibegovic:2017} with a negligible amount of crystal defects, and the development of in-situ growth of epitaxial semiconductor/superconductor hybrids\cite{Krogstrup:2015} enables reproducible, uniform contacting and improved character of induced superconductivity\cite{Chang:2015}. Finally, in-situ shadow patterning methods have been developed to  reduce post growth processing\cite{Carrad:2020,Khan2020,gazibegovic:2017,Krizek:2017,Heedt:2020} and thus significantly reduce random scattering from surface adsorbents, and improve reproducibility and electrostatic stability. 
With minimal disorder and extrinsic random scattering, intrinsic properties such as the non-uniform carrier distribution can be dominating. In low band gap III-V-compounds such as InAs or InSb the surface Fermi level is pinned in the conduction band\cite{Olsson:1996} and for the NW geometry, the expectation is a cylindrical electron accumulation layer as also found in Schrödinger-Poisson modeling\cite{winkler_unified_2019,Mikkelsen:2018, Antipov:2018}. Early Kelvin probe microscopy\cite{halpern_direct_2012} and studies of the scaling of room-temperature resistance with the diameter\cite{dayeh_transport_2009, Scheffler:2009} is consistent with this scenario, however, clear confirmation in quantum devices operating at low temperature is lacking. 

Quantum interference in Josephson devices in a magnetic field provides a sensitive probe of the spatial electron distribution and has been used to study the emergence of edge states in topological insulators\cite{Hart:2014}. Here we measured the magnetoconductance of high mobility InAsSb NWs with epitaxial aluminum grown on two facets of the hexagonal nanowire. We consider both devices with a continuous Al and devices where a Josephson junction (JJ) is realized by in-situ shadow lithography\cite{Khan2020}. Interference patterns periodically modulated by the flux through the NW cross-section are reproducibly found over all six devices and confirm the presence of a proximitized cylindrical accumulation layer at the NW surface. We show that this property is not limited to the JJ geometry but relevant also in uniform hybrid NWs and in both classes of devices, electrostatic gating tunes the density and spatial distribution of the carriers and the cylindrical geometry is lost at large negative gate potentials consistent with electrostatic modeling\cite{winkler_unified_2019,Mikkelsen:2018, Antipov:2018}. The results are important for the interpretation result obtained from NW quantum devices and we discuss the relation to the Little-Parks effect observed in full shell hybrid NWs \cite{vaitiekenas:2020}.\\

\begin{figure*}
\includegraphics[width=\textwidth]{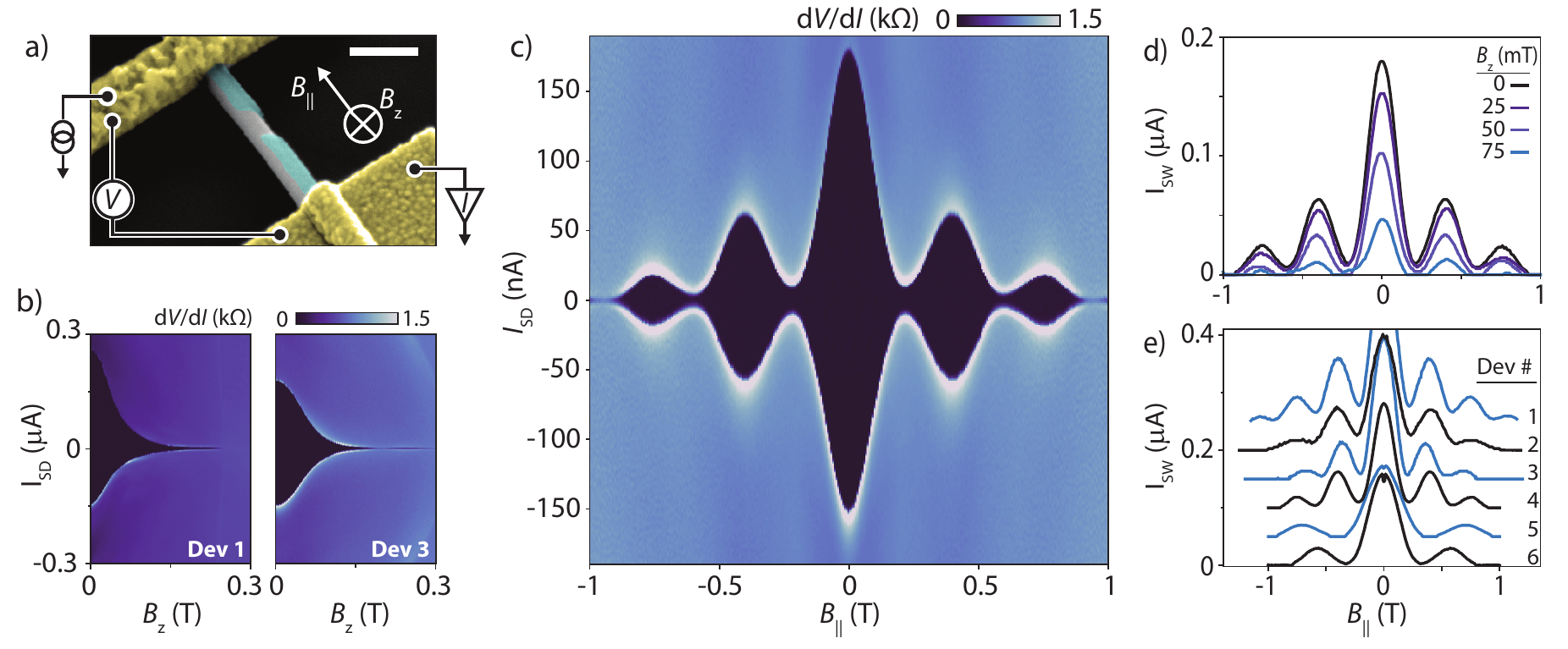}
\caption{\label{fig:Fig2} (a) Scanning electron micrograph of a Josephson junction (JJ) device; scale bar: 250 nm. (b) $\dVdI$ for two JJ devices as a function of $I$ and $B_\mathrm{z}$. (c) $\dVdI$ vs.\ $B_{\parallel}$ for Dev.\ 3 exhibiting pronounced periodic modulation of $I_\mathrm{sw}$. (d) Switching current vs.\ $B_{\parallel}$ for fixed values of $B_\mathrm{z}$. The modulation remains with a uniform decrease in amplitude. (e) Switching current vs $B_{\parallel}$ for all measured JJ devices confirming reproducibility.}
\end{figure*} 

\subsection{Experimental details}
Devices were fabricated using InAsSb NW grown from (111)B facets etched into a (100) InAs substrate\cite{Khan2020,gazibegovic:2017}. The InAsSb ternary alloy was chosen as a compromise between the high mobility and low effective mass of InSb and the higher chemical robustness of InAs\cite{Khan2020, Sestoft:2018}. The NWs feature a hexagonal cross section and following semiconductor growth,  $\sim 15\, \mathrm{nm}$ of epitaxial aluminum was grown on two of the 6 facets without breaking the reactor vacuum.\cite{Krogstrup:2015, Khan2020} For JJ devices, $\sim 100 \, \mathrm{nm}$ wide gaps in the Al were realized in-situ by shadowing Al growth by adjacent NWs\cite{Krizek:2017, gazibegovic:2017, Khan2020}. The obviation of subsequent fabrication steps and etching leads to low disorder and electrostatically stable devices\cite{Carrad:2020}. For measurements, individual NWs were transferred to standard Si/SiO$_\mathrm{x}$ substrates where the doped Si acts as back-gate $(V_\mathrm{BG})$ isolated from the NWs by $200 \, \mathrm{nm}$ of $\mathrm{SiO}_2$. Ti/Au Ohmic contacts were defined by standard lithography and Figs.\ \ref{fig:Fig2}a and \ref{fig:Fig1}a show scanning electron micrographs of a finished device with shadow junctions and continuous Al, respectively.
While transferring NWs to the device substrate, the orientation of the Al facets to the substrate plane was not controlled, and devices with a significantly reduced gate-response were attributed to hybrids with Al facing the back-gate. Such samples were not included in further measurements. Measurements were performed in a dilution refrigerator with a base temperature of $15 \, \mathrm{mK}$ and equipped with a 3D vector magnet. For each device the orientation of the nanowire axis was determined with an accuracy of $\lesssim 2 ^\circ$ by measuring the anisotropy of the critical magnetic field and this allowed application of magnetic field $(B_\parallel )$ along the axis of the nanowires. For uniform InAsSb/Al devices the differential resistance $\mathrm d V / \mathrm d I$ was measured in a four-terminal configuration (Fig.\ \ref{fig:Fig1}a) while JJ segments were measured in a pseudo four-terminal configuration and contact resistances, identified as the baseline resistance in the superconducting state, were subsequently subtracted. In total, nine devices were measured; six with JJs (Dev.\ 1-6) and three without (Dev.\ 7-9). In each category all devices showed qualitatively consistent behavior and in the following, results from one representative device of each type is presented; results from remaining devices are included in the Supporting Information.

\begin{figure*}
\includegraphics[width=\textwidth]{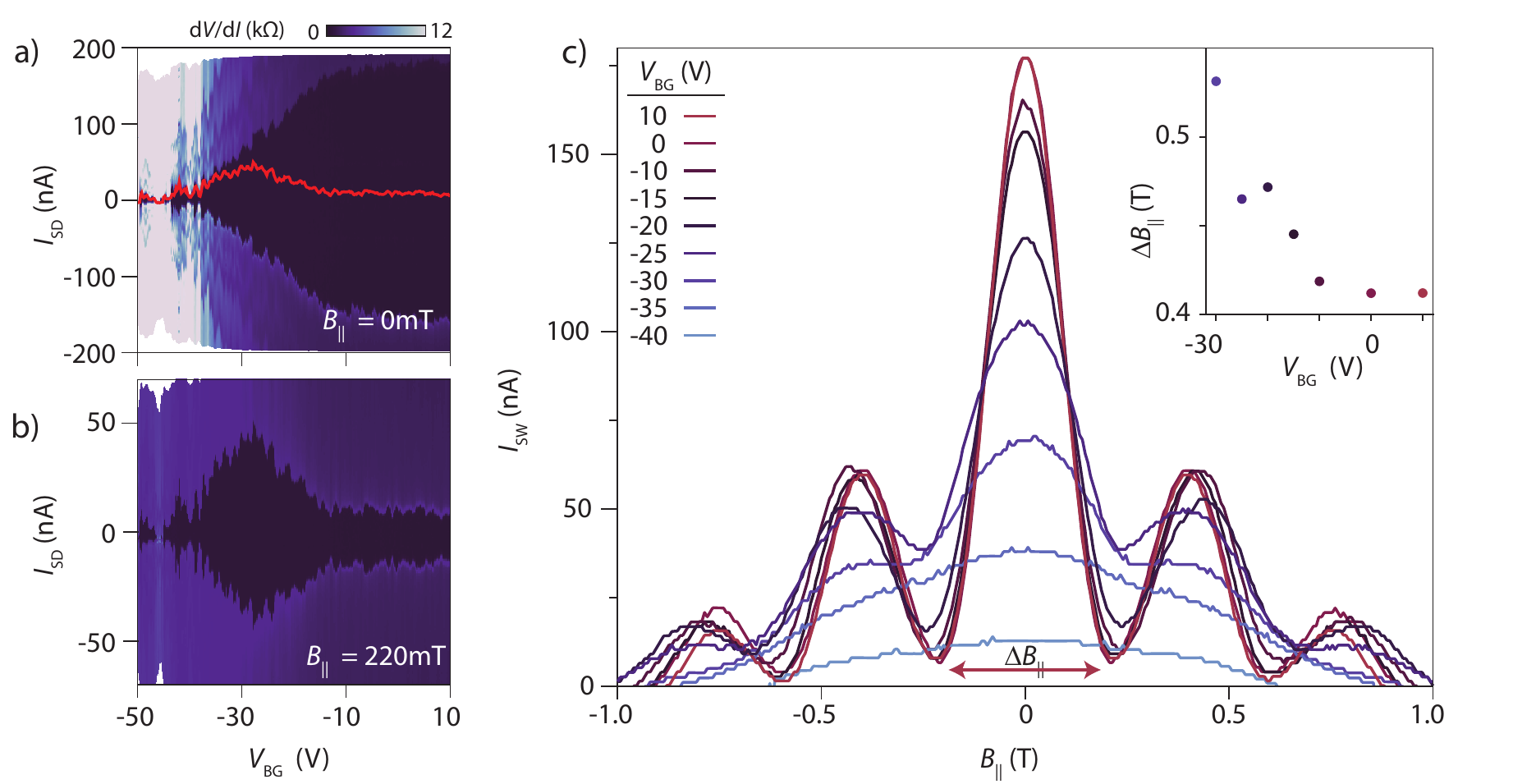}
\caption{\label{fig:Fig4} (a) Differential resistance vs.\ $I_\mathrm{DC}$ and $V_\mathrm{BG}$ for $B=0\, \mathrm T$. Sweep direction from negative to positive. The red line indicates $I_\mathrm{sw}(B_\parallel = 220 \, \mathrm{mT})$ extracted from (b). (b) As (a) for $B_\parallel = 220 \, \mathrm{mT}$, showing the non monotonic $V_\mathrm{BG}$-dependence due to interference. (c) $I_{sw}$ vs.\ $B_{||}$ at different $V_\mathrm{BG}$ showing a monotonous decrease of peak values while minima increase. The inset shows the $V_\mathrm{BG}$-dependence of the oscillation period (see text).}
\end{figure*}

\subsection{Results}
We consider first the JJ devices which isolates the contribution of the semiconductor and then subsequently discuss the situation in the uniform hybrid nanowires with continious Al. Figure \ref{fig:Fig2}b shows the differential resistance $\dVdI$ of JJ Devices \ 1 and 3 as a function of the perpendicular out-of-plane magnetic field $B_{\mathrm{z}}$ sweeping $I$ from negative to positive. A zero-resistance superconducting region is observed at low currents and the switching current where the device switches to a finite resistance state decreases monotonously with $B_\mathrm{z}$ and at $B_\mathrm{z}^\mathrm c \sim 150 \, \mathrm{mT}$ superconductivity is completely suppressed consistent with previous measurements of epitaxial hybrid nanowires in a perpendicular field \cite{Chang:2015,Heedt:2020}. A strikingly different behavior is observed in the corresponding measurement performed with the field $B_\parallel$ applied parallel to the NW (Fig.\ \ref{fig:Fig1}c). Pronounced oscillations of the switching current appear with a period of $\sim 440 \, \mathrm{mT}$. The amplitude decreases towards higher $B_\parallel$ as superconductivity is gradually suppressed but due to the thin Al, the critical magnetic field is larger in the parallel direction and although it is strongly suppressed at the minima of the oscillations, the switching current remains finite up to the highest applied magnetic field of 1 T. In Fig.\ \ref{fig:Fig2}d the dependence of the switching current on $B_\parallel$ has been extracted from measurements similar to Fig.\ \ref{fig:Fig2}c repeated for various $B_\mathrm{z}$. The oscillation amplitude decreases with $B_\mathrm{z}$ following the envelope in Fig.\ \ref{fig:Fig2}b, however the period and qualitative shape of the $I_\mathrm{sw}$ modulation is independent of $B_\mathrm{z}$. The experiment was repeated for a number of devices and Fig.\ \ref{fig:Fig2}e shows the switching current $I_\mathrm{sw}(B_\parallel)$ for 6 JJ devices extracted from full maps like Fig.\ \ref{fig:Fig2}c; (see SI Fig.\ S4). All devices exhibit qualitatively similar periodic modulation with a small variation in the period between the different devices as discussed below.\\

\paragraph{Gate dependence of interference pattern} 
The $I_\mathrm{sw}$ modulation in Fig.\ \ref{fig:Fig2}c is the main experimental result of this work and we now consider the effect of changing the overall carrier density in the junction using the back-gate. Figure \ref{fig:Fig4}a shows the differential resistance vs.\  $V_\mathrm{BG}$ and $I_\mathrm{SD}$ for $B=0 \, \mathrm{mT}$. The switching current appears as the first contour with a non-zero $\dVdI$ and decreases from $\sim 200 \, \mathrm{nA}$ to zero upon lowering $V_\mathrm{BG}$ from $10 \, \mathrm V$ to $-50 \, \mathrm V$ because the $n$-type semiconductor is depleted of carriers\cite{Doh:2005, Khan2020}. In Fig.\ \ref{fig:Fig4}c, the dependence of $I_\mathrm{sw}$ on $B_\parallel$ is shown for different $V_\mathrm{BG}$ (see SI Fig.\ S6 for full maps). The periodic nature of $I_\mathrm{sw}$-modulations is not affected by $V_\mathrm{BG}$. However as the junction is depleted the peak amplitude decreases and the  period of the modulation increases slightly as emphasized in the inset, which shows the $V_\mathrm{BG}$ dependence of the difference $\Delta B_\parallel$ between the two first minima as indicated on the Fig.\ \ref{fig:Fig4}c (SI Fig.\ S11). Only close to pinch-off does the modulation disappear. 

At the minima of the oscillations, the switching current increases with decreasing $V_\mathrm{BG}$ despite the increasing NW resistance. This unusual behavior is emphasized in Fig.\ \ref{fig:Fig4}b which shows a measurement similar to Fig.\ \ref{fig:Fig4}a, however, measured with $B_\parallel = 220 \, \mathrm{mT}$ close to the first minimum. At $V_\mathrm{BG} = 10 \, \mathrm V$ a highly suppressed $I_\mathrm{SW} \sim 10 \, \mathrm{nA}$ is observed. Upon decreasing $V_\mathrm{BG}$, $I_\mathrm{sw}$ remains constant until $V_\mathrm{BG} \sim -15 \, \mathrm{V}$ from where on $I_\mathrm{SW}$ first increases as the interference pattern is gradually lost, to a maximum of $\sim 40 \, \mathrm{nA}$ at $V_\mathrm{BG} \sim -30 \, \mathrm V$ after which it re-traces the decrease observed for $B_\parallel = 0$. For comparison, this unconventional $I_\mathrm{SW}(V_\mathrm{BG})$ for $B_\parallel = 220 \, \mathrm{mT}$ is extracted and shown in Fig.\ \ref{fig:Fig4}(a) (red trace).\\

\begin{figure}
\includegraphics[width=8.5cm]{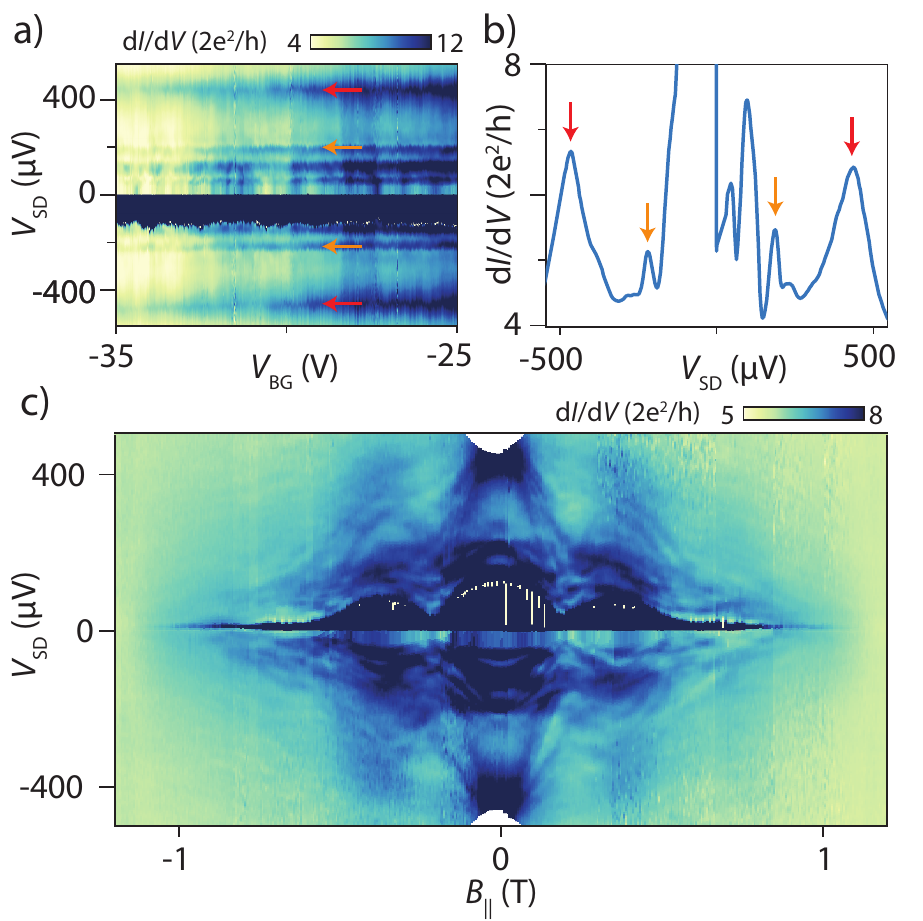}
\caption{\label{fig:Fig5} (a) $dI/dV$ vs.\ $V_\mathrm{SD}$ and $V_\mathrm{BG}$ of Dev.\ 2 for at $B_{||} = 0$. Arrows indicate peaks due to multiple Andreev reflection appearing independent of $V_\mathrm{BG}$ . (b) Line cut of (a) at $V_\mathrm{BG} = -30V$. (c) The evolution of the line-cut in b) while varying $B_{||} = 0$ showing periodically modulated Andreev Bound States.
 }
\end{figure}

Fraunhofer interference of supercurrents is a typical source of $I_\mathrm{sw}$-modulations in JJs due to field induced spatial dependence of the superconducting phase in the leads. Therefore, two different measurements were performed to investigate whether the observed effect is specific to the JJ geometry. First, by operating the JJ-devices with $V_\mathrm{BG}$ close to pinch-off, measurements at finite voltage bias $V_\mathrm{SD}$ effectively performs spectroscopy of the density of states in the leads. Secondly, direct measurements of the effect of $B_\parallel$ on uniformely coated nanowires without JJ were performed.

\paragraph{Spectroscopy at finite bias} 
Figure \ref{fig:Fig5}a shows $\dVdI$ as a function of the DC bias voltage $V_\mathrm{SD}$ and $V_\mathrm{BG}$. The sweeping direction is from positive to negative. Heavy filtering of the DC lines makes the measurement effectively current-biased at low $V_\mathrm{SD}$ and a smooth re-trapping transition compared to the abrupt switching-current results in the asymmetry of the low-bias part of the figure. A series of symmetric features are observed (arrows) at voltages which do not depend on the junction transparency (which decreases with $V_\mathrm{BG}$) and are therefore associated with properties of the leads. The features are also emphasized in panel (b) showing a trace extracted for fixed $V_\mathrm{BG} = -30 \, \mathrm V$ and follow the expected harmonic sub gap structure due to multiple Andreev reflections (MAR) as commonly observed in nanowire JJs \cite{Doh:2005,ridderbos_multiple_2019, shailos_proximity_2007,Nilsson:2012}. The extracted gap is $\Delta = 225 \, \mu \mathrm{eV}$ consistent with the expected gap of the Al hybrids \cite{Chang:2015}. In addition to the MAR features, spectroscopic measurements  also probe sub-gap states in the leads\cite{Heedt:2020} and Fig.\ \ref{fig:Fig5}c shows a measurement of the spectrum as a function of $B_\parallel$ for a fixed $V_\mathrm{BG} = -30 \, \mathrm{V}$. A rich spectrum emerges at finite $B_\parallel$ which is originating from the $2\Delta$ peak and is following a pattern modulating similar to the switching current in Figs.\ 1-2. This demonstrates that the observed interference is not exclusively a junction property. We attribute the fainter features in Fig.\ \ref{fig:Fig5}c which coincide at $B_\parallel = 0$ with $2\Delta/e$ and $\Delta/e$ and smoothly closes around $\pm 1 \, \mathrm T$ to the sub-gap-structure reflecting the slow dependence of $\Delta$ on $B_\parallel$.

\begin{figure}
\includegraphics[width=8.5 cm]{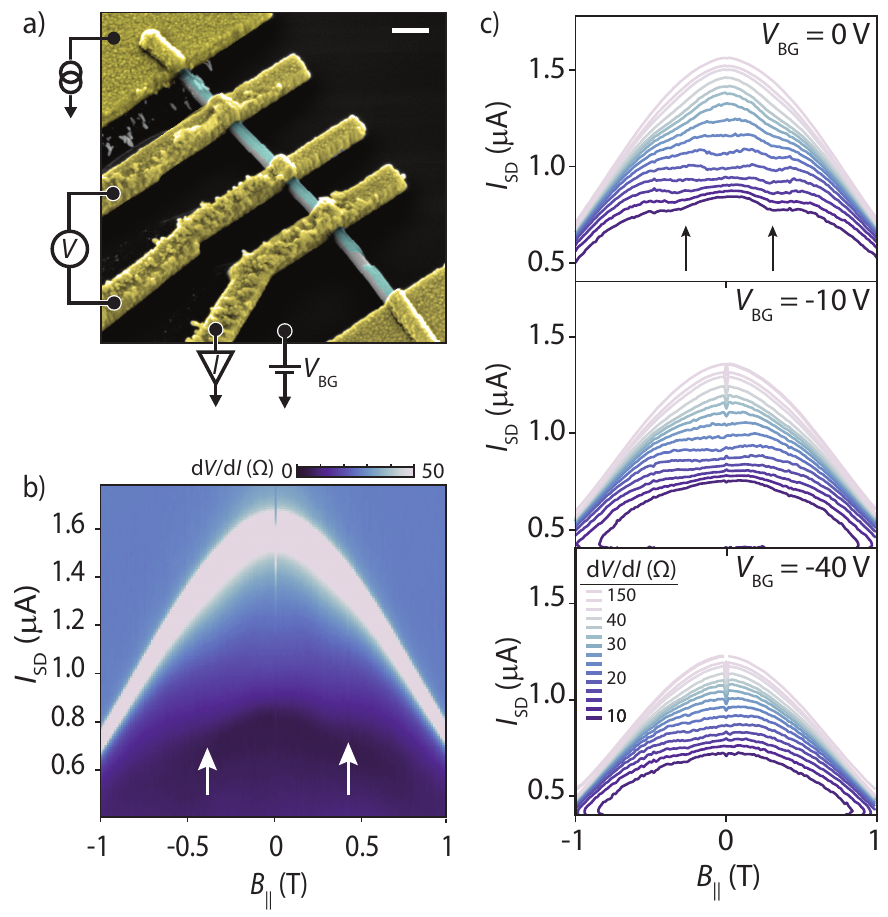}
\caption{\label{fig:Fig1} (a) False colour SEM micrograph of a InAsSb device. Epitaxial Al is shaded blue with a shadow gap  in the lower segment (JJ). Scale bar is $250 \, \mathrm{nm}$. (b) Color map of $\dVdI$ vs.\ $I$ and $B_\parallel$ at $V_\mathrm{BG} = 10 V$ and $T=20 \mathrm{mK}$. Arrows indicate kinks in the otherwise monotonic decrease of critical current with $B_\parallel$. (c) The Contours from measurement similar to (b) at different $V_\mathrm{BG}$ in steps of $10, 12.5, 15,...,35,40,50,60,150 \Omega$. Arrows highlight the oscillatory structure which disappears with decreasing gate.}
\end{figure}

\paragraph{Uniform nanowires} 
As a final confirmation we consider the measurements of devices with continuous Al (Fig.\ \ref{fig:Fig1}a). Due to the higher charge carrier density, transport is dominated by the Al, however, as the metal shell is not susceptible to gating, the contribution of the semiconductor can be identified by the dependence on $V_\mathrm{BG}$. 
Figure \ref{fig:Fig1}b shows the dependence of four terminal $\dVdI$ on $I$ and $B_\parallel$. The switching current appears as a pronounced increase in the resistance (white contour), and the finite resistance observed also for $I < I_\mathrm{sw}$ is attributed to phase slips and is commonly observed in 1D superconductor NWs \cite{li_switching_2011, golubev_anomalous_2018, rogachev_influence_2005, sahu_individual_2009}. The zero field $I_\mathrm{sw} \sim 1.6 \, \mu \mathrm A$ and the critical magnetic field $B^c_\parallel \sim 1.6 \, \mathrm T$ (extrapolated from Fig.\ \ref{fig:Fig1}b) is similar but slightly lower than previous studies of epitaxial hybrid nanowires\cite{Meservey:1972,Krogstrup:2015}, which might be due to a grainy Al. The switching current decreases with $B_\parallel$ and exhibits kinks at $B_\parallel \approx 0.4 \, \mathrm{T}$ (arrows) which are more clearly seen in Fig.\ \ref{fig:Fig1}c which shows constant-resistance contours extracted from Fig.\ \ref{fig:Fig1}c and corresponding measurements at three different $V_\mathrm{BG}$. The zero field switching current decreases from $1.6 \mu \mathrm A$ for $V_\mathrm{BG} = 0 \, \mathrm V$ to $1.1 \mu \mathrm A$ for $V_\mathrm{BG} = -40 \, \mathrm V$, and the kinks in the contours disappear upon lowering $V_\mathrm{BG}$. The gate dependence is consistent with a supercurrent carried partly by the Al shell and partly by the semiconductor. Here, the latter contribution decreases as the semiconductor is depleted at low $V_\mathrm{BG}$ and since the kinks at finite $B_\parallel$ also disappear these are related to the presence of the semiconductor. The qualitative resemblance of these results with the JJ measurements serves as an additional indication that the interference is not a JJ effect but a property of the hybrid nanowire materials.

\section{Discussion}
The regular $I_\mathrm{sw}$ modulation observed here -- i.e. fringes with regular period in $B$ that behave monotonically with $V_\mathrm{BG}$ -- contrasts previous reports on supercurrent interference in NW devices, which generally feature aperiodic patterns in field and gate voltage due to random interference\cite{Zuo:2017,Gharavi:2015,Gharavi:2017}. Depending on the current distribution, conventional Fraunhofer interference can lead to periodic oscillations\cite{Hart:2014,Paajaste:2015}, however, such a mechanism is inconsistent with the observed modulation in the JJ leads (Fig.\ \ref{fig:Fig5}) and in the junctionless devices (Fig.\ \ref{fig:Fig1}). A key feature of the observed behaviour is that only the field component parallel to the NW modulates $I_\mathrm{sw}$, suggesting a relation to states acquiring a phase while encircling the NW surface. This is also consistent with the period of the oscillations, which we associate with the field required to add a flux quantum $\Phi_0$ through a typical cross-sectional area. Assuming a cylindrical geometry we get $\Delta B = h/2 \pi e (d_\phi/2)^2$, where $d_\phi$ is the effective diameter, $h$ is the Planck constant and $e$ the electron charge. For devices 1-4, $\Delta B$ corresponds to a diameter $d_\Phi = 80-85$~nm, and for devices 5 and 6, the corresponding flux diameter was $d_\Phi=60-65$~nm. The physical nanowire diameters, $d_\mathrm{NW}$, were independently estimated from SEM micrographs (Supporting Fig. S1) as $\sim 120$~nm and $\sim 90$~nm for devices 1-4 and 5-6, respectively. Given that numerical and theoretical estimates have shown the surface accumulation layer thickness to be 15-20 nm, the period $\Delta B$ is consistent with electron interference in the accumulation layer. A main result of the current work is therefore that transport in the InAsSb NW devices is dominated by carriers occupying a surface accumulation layer.

We now consider the physical mechanism leading to the $I_\mathrm{sw}$ behaviour. In an Aharonov-Bohm (AB) ring between two superconductors with phase difference $\varphi$, Andreev bound states (ABS) form with energy\cite{Dolcini:2007} 
\begin{equation}
E(\Phi) = \pm \Delta \sqrt{1-\tau(\Phi) \sin^2((\varphi + \theta(\Phi))/2)}
\end{equation}
Here, $\theta = 2\pi \Phi / \Phi_0$ is the AB phase, $\Phi$ is the flux penetrating the ring, and $\tau (\Phi)$ is the flux-dependent transmission. Extending this concept to the situation in the JJ leads -- which consist of a hybrid NW partially covered by the Al film -- ABS form around the NW circumference due to Andreev reflection from the single superconductor, corresponding to $\varphi=0$. This situation is similar to reflectionless tunneling\cite{vanWees:1992}, albeit with a fixed area for phase accumulation here. The result is a periodic modulation of the semiconductor/superconductor coupling, as previously observed for nanoscale core/shell cylinders\cite{Gul:2014}. The ABS energies are modulated by $\theta(\Phi)$ and are minimal at $\theta = \pi$. For a diameter of 80~nm, the first minima would be expected to occur at $B=210 \, \mathrm{mT}$,\cite{Zhou:1998,le_Sueur:2008} consistent with the spectroscopic measurements in Fig.\ \ref{fig:Fig5}. The modulation of the $I_\mathrm{sw}$ in the uniform hybrid nanowires in Fig.\ \ref{fig:Fig1} can thereby be accounted for by the periodic modulation of the ABS energies, combined with a reduction of the InAsSb/Al coupling.

In the case of the JJ devices, for $I<I_\mathrm{sw}$, the phase difference across the junction, $\varphi$, varies with $I$ to ensure a dissipation-less supercurrent $ \propto \mathrm d E/ \mathrm d \varphi$. Therefore, an AB-phase $\theta(\Phi)$ only acts as an offset, which can be compensated by $\varphi$ and thus not result in $I_\mathrm{sw}$ modulations. Instead, the observed $I_\mathrm{sw}$ oscillations in the JJs are a consequence of the transmission $\tau(\Phi)$, which oscillates due to the conventional AB effect \cite{aharonov_significance_1959}. 
We note that resistance oscillations were not clearly resolved in the normal state for $T>T_\mathrm{c}$ which we attribute to a shorter phase coherence length at elevated temperature or the dominating contribution of channels with high transmission in the superconducting state\cite{Murani:2017}. 

The number of transverse modes and their degeneracies in the hybrid nanowires, as well as the gatability and coupling to magnetic fields, are important for interpreting results and the prospects of realizing theoretical proposals for topological devices\cite{Oreg:2010,Lutchyn:2010,vaitiekenas:2020}. The gate-dependence of the interference pattern in Fig.\ 4 probes the changes of the density profile. Due to screening by the grounded Al half shell, the gate acts more strongly on the JJ segment than below the superconductor, and affects most strongly the part of this segment facing the back gate. At $V_\mathrm{BG} = 0 \, \mathrm{V}$ the interference pattern and significant suppression of $I_\mathrm{sw}$ at $\Phi = \Phi_0/2$ shows that electrons occupy the cylindrical accumulation layer near the surface of the NW. Upon decreasing $V_\mathrm{BG}$ the resistance increases and $I_\mathrm{sw}(B_\parallel = 0)$ decreases as carriers are depleted from the junction. The interference pattern remains qualitatively the same, showing that a accumulation layer persists. The slight increase of the period $\Delta B_\parallel$ with decreasing $V_\mathrm{BG}$ seen in Fig.\ \ref{fig:Fig4}(c) was also observed for the other devices (SI Fig.\ S7-8) and is attributed to a gradual shift of the carrier distribution towards the core of the NW. Reducing the effective area of the carriers in this way results in a larger $\Delta B_\parallel$, and ultimately causes the loss of the cylindrical geometry. At the lowest $V_\mathrm{BG}$ the junction acts as a high resistance weak link, with almost no $I_\mathrm{sw}$ oscillation (see Fig.\ \ref{fig:Fig4}c). However, flux modulation remains in the finite bias spectroscopy in Fig.\ \ref{fig:Fig5} and Supporting Figure S9 and S10, showing that the presence of the cylindrical accumulation in the uniform half-shell hybrid segments constituting the leads of the junction remains, even when the junction is completely depleted. 

Finally, we comment on the relevance to recent reports of topological superconductivity in full-shell hybrid nanowires\cite{vaitiekenas:2020}. For our devices, at a finite $B_\parallel$ the Andreev pairs undergo a phase winding upon encircling the NW similar to superconducting cylinders in the Little-parks regime of Ref.~\citenum{vaitiekenas:2020}. However, the surface accumulation layer demonstrated in this work does not host a phase coherent condensate as in the Little-Parks scenario \cite{little_observation_1962}. Therefore, coupling to states of the NW core\cite{Island:2017} is not expected to produce topological states,\cite{vaitiekenas:2020} which requires at least 3 different superconducting phases \cite{Lesser:2020}. Indeed, the Shapiro step measurements in Supporting Fig.\ S11 show no sign of a change in the periodicity of the current phase relation at finite $B_\parallel$, which is the predicted hallmark of a topological phase transition \cite{kwon_fractional_2003}.

In conclusion, we have presented a detailed study of the superconducting properties of low disorder InAsSb/Al epitaxial hybrid uniform nanowires and Josephson junctions realized by shadow lithography.\cite{Carrad:2020, Khan2020, gazibegovic:2017} A pronounced interference pattern of the switching current is observed in both cases when applying a magnetic field parallel to the nanowire axis.
The results were explained by Andreev bound states encircling the nanowire, subject to a phase winding due to applied magnetic flux. The oscillation period in $B_\parallel$ matches that expected from a surface accumulation layer of thickness $15-20$~nm \cite{winkler_unified_2019,Mikkelsen:2018, Antipov:2018}. Electrostatic gating facilitated alteration -- and ultimate loss -- of the cylindrical accumulation layer geometry.

The results are repeated for many different devices, and we suggest that compared to earlier experiments,\cite{Zuo:2017} the effects of the surface accumulation layer observed here is related to the high mobility of the InAsSb \cite{thelander_electrical_2012}, combined with the reduction of impurity scattering and disorder due to epitaxial interfaces and shadow lithography. 
We therefore expect these effects will be of increasing importance as materials and device quality continues to improve, and the results thus hold importance for interpretation of results in superconducting quantum devices with hybrid nanowires.

\textbf{Acknowledgement} We thank Karsten Flensberg for useful discussions. This research was supported by the Danish National Research Foundation. S.A.K, and P.K. were funded by European Union Horizon 2020 research and innovation program under the Marie Sk\l{}odowska-Curie Grant No. 722176 (INDEED), Microsoft Quantum and the European Research Council (ERC) under Grant No. 716655 (HEMs-DAM). T.S.J was supported by research grants from Villum Fonden (00013157), The Danish Council for Independent Research (7014-00132), and European Research Council (866158). 

\subsection*{Supporting Information} The supporting information contains scanning electron micrographs and data from all six devices studied, and is available at \url{https://sid.erda.dk/share_redirect/EXlG3QDvfM}
\bibliography{refs,PapersThesis,additionalARref}

\end{document}